\documentclass[a4paper,12pt]{article}
\usepackage{mathrsfs}
\usepackage{bbm}
\usepackage{epsf,amsmath,amssymb}
\usepackage[dvips,usenames]{color}
\usepackage{graphicx}
\usepackage{color}
\usepackage{colortbl}

\newlength{\dinwidth}
\newlength{\dinmargin}
\def\pslash{\rlap{\hspace{0.02cm}/}{p}}
\def\kslash{\rlap{\hspace{0.02cm}/}{k}}
\newcommand{\half}{\frac{1}{2}}
\begin{document}
\title{\bf Constraints on the nonuniversal $Z^\prime$ couplings from
$B\to\pi K$, $\pi K^{\ast}$ and $\rho K$ Decays}
\author{Qin Chang$^{a,b}$, Xin-Qiang
Li$^{c}$\footnote{Alexander-von-Humboldt Fellow}, Ya-Dong Yang$^{a,d}$
\footnote{Corresponding author}\\
{ $^{a}$\small Institute of Particle Physics, Huazhong Normal
University, Wuhan, Hubei  430079, P.~R. China}\\
{ $^b$\small Department of Physics, Henan Normal University,
Xinxiang, Henan 453007, P.~R. China}\\
{ $^c$\small Institut f\"ur Theoretische Physik E, RWTH Aachen
University, D--52056 Aachen, Germany}\\
{ $^d$\small Key Laboratory of Quark \& Lepton Physics, Ministry of
Education, P.~R. China}}
\date{}
\maketitle
\bigskip\bigskip
\maketitle \vspace{-1.5cm}

\begin{abstract}
Motivated by the large difference between the direct CP asymmetries
$A_{CP}(B^-\to \pi^0 K^-)$ and $A_{CP}(\bar{B}^{0}\to \pi^{+}
K^{-})$,  we combine the up-to-date experimental information on
$B\to\pi K$, $\pi K^{\ast}$ and $\rho K$ decays to pursue possible
solutions with the nonuniversal $Z^{\prime}$ model. Detailed
analyses of the relative impacts of different types of couplings are
presented in four specific cases. Numerically, we find that the new
coupling parameters, $\xi^{LL}$ and $\xi^{LR}$ with a common
nontrivial new weak phase $\phi_L\sim-86^{\circ}$, which are
relevant to the $Z^{\prime}$ contributions to the electroweak
penguin sector $\triangle C_9$ and $\triangle C_7$, are crucial to
the observed ``$\pi K$ puzzle". Furthermore, they are found to be
definitely unequal and opposite in sign. We also find that
$A_{CP}(B^-\to \rho^0 K^-)$ can put a strong constraint on the new
$Z^{\prime}$ couplings, which implies the $Z^{\prime}$ contributions
to the coefficient of QCD penguins operator $O_3$ involving the
parameter $\zeta^{LL}$ required.

\end{abstract}

\newpage

\section{Introduction}
During the past several years, the observed discrepancies between
the experimental measurements and the theoretical predications
within the Standard Model~(SM) for several observables in $B\to \pi
K$ decays, the so-called ``$\pi K$ puzzle"~\cite{pikpuz}, have
attracted much attention. Extensive investigations both within the
SM~\cite{Beneke2,Beneke3,YD,PikPQCD,PikSCET,Pik}, as well as with
various specific New Physics~(NP)
scenarios~\cite{YDYang,Barger,PikNP}, have been performed.

Averaging the recent experimental data from BABAR~\cite{bar},
Belle~\cite{Belle_Nature}, CLEO~\cite{cleocp} and CDF~\cite{cdfcp},
the Heavy Flavor Averaging  Group~(HFAG) gives the following
up-to-date rsults~\cite{HFAG}
\begin{eqnarray}\label{AcpPuzzleexp}
 &&A_{CP}(B^{-}\to K^{-}\pi^{0})=0.050\pm 0.025~,\nonumber\\
 &&A_{CP}(\bar{B}^{0}\to K^{-}\pi^{+})=-0.097\pm 0.012,
 \end{eqnarray}
from which the difference between direct CP violations in the
charged and the neutral modes
\begin{eqnarray}\label{cpdiff}
\Delta A \equiv A_{CP}(B^{-}\to K^{-}\pi^{0})- A_{CP}(\bar{B}^{0}\to
K^{-}\pi^{+})=0.147\pm0.028
\end{eqnarray}
is now established at about $5\sigma$ level.

Theoretically, it is generally expected that within the SM, these
two CP asymmetries $A_{CP}(\bar{B}^0_d\to\pi^+ K^-) $ and
$A_{CP}(B^-_u\to\pi^0 K^-)$ should be approximately equal. For
example, based on the QCD factorization
approach~(QCDF)~\cite{Beneke1}, the recent theoretical predictions
with two different schemes for the end-point divergence are
\begin{eqnarray}
\label{AcpSchemeI} &&\left\{\begin{array}{l}
A_{CP}(B^-_u\to\pi^0 K^-)=-3.6\%~, \\
A_{CP}(\bar{B}^0_d\to\pi^+ K^-)=-4.1\%~,
\end{array}\right.~{\rm Scheme~I~(Scenario~S4)}~\cite{Beneke3}\,,\\
\label{AcpSChemeII} &&\left\{\begin{array}{l}
A_{CP}(B^-_u\to\pi^0 K^-)=-10.8\%~, \\
A_{CP}(\bar{B}^0_d\to\pi^+ K^-)=-12.4\%~,
\end{array}\right.~{\rm Scheme~II}~(m_g=0.5~{\rm MeV})~\cite{YDYang}\,.
\end{eqnarray}
Here, the Scheme~I is the way to parameterize the end-point
divergence appearing in hard-spectator and annihilation corrections,
by complex parameters $X_{A,H}=\int^{1}_{0}
dy/y=\mathrm{ln}(m_b/\Lambda) (1+\rho_{A,H} e^{i\phi_{A,H}})$, with
$\rho_{A,H} \leq 1$ and unrestricted $\phi_{A,H}$~\cite{Beneke3}.
The Scheme~II, as an alternative to the first one, is the way to
quote the infrared finite gluon propagator to regulate the
divergence. It is interesting to note that an infrared finite
behavior of gluon propagator are not only obtained by solving the
well-known Schwinger-Dyson equation~\cite{Cornwall,Aguilar,Alkofer},
but also supported by recent Lattice QCD simulations~\cite{lattice}.
However, both of these two schemes suffer the mismatch of $\Delta A$
given by Eq.~(\ref{cpdiff}). Furthermore, within the framework of
perturbative QCD approach~(pQCD)~\cite{KLS}, and the soft-collinear
effective theory~(SCET)~\cite{scet}, the theoretical predictions
read
\begin{eqnarray}
\label{AcpPuzzlePQCD} &&\left\{\begin{array}{l}
A_{CP}(B^-_u\to\pi^0 K^-)_{PQCD}=(-1^{+3}_{-5})\%~, \\
A_{CP}(\bar{B}^0_d\to\pi^+ K^-)_{PQCD}=(-9^{+6}_{-8})\%~,
\end{array}\right.~{\rm pQCD}~\cite{PikPQCD}\,,\\
\label{AcpPuzzleSCET} &&\left\{\begin{array}{l}
A_{CP}(B^-_u\to\pi^0 K^-)_{SCET}=(-11\pm9\pm11\pm2)\%~, \\
A_{CP}(\bar{B}^0_d\to\pi^+ K^-)_{SCET}=(-6\pm5\pm6\pm2)\%.
\end{array}\right.~{\rm SCET}~\cite{PikSCET}\,.
\end{eqnarray}
Obviously, the present theoretical estimations within the SM are not
consistent with the established $\Delta A$. The mismatch might be
due to our current limited understanding of the strong dynamics
involved in hadronic B decays, but equally also to possible NP
effects~\cite{peskin,mannel}.

In some well-motivated extensions of the SM, additional
$U(1)^{\prime}$ gauge symmetries and associated $Z^{\prime}$ gauge
boson could arise. Searching for the extra $Z^{\prime}$ boson is an
important mission in the experimental programs of
Tevatron~\cite{Tevatron} and LHC~\cite{LHC}. Performing the
constraints on the new $Z^{\prime}$ couplings through low-energy
physics, on the other hand, is very imporatnt for the direct
searches and understanding its phenomenology. Theoretically, the
flavor changing neutral current (FCNC) is forbidden at tree level in
the SM. One of the simple extensions is the family nonuniversal
$Z^{\prime}$ model, which could be naturally derived in certain
string constructions\cite{string}, $E_6$ models\cite{E6} and so on.
It is interesting to note that the nonuniversal $Z^{\prime}$
couplings could lead to FCNC and new CP-violating
effect~\cite{Langacker}, which possibly provide a solution to the
afore mentioned ``$\pi K$ puzzle". With some simplifications of the
nonuniversal $Z^{\prime}$ model and neglecting the color-suppressed
electroweak~(EW) penguins and the annihilation amplitudes,
Ref.~\cite{Barger} gets four possible solutions
\begin{eqnarray} \label{Barger}
A_{L}:  &&\{\xi^{LL},\phi_L\}= \{0.0055,110^{\circ}\}\,,
 \qquad  B_{L}:  \{\xi^{LL},\phi_L\}= \{0.0098,-97^{\circ}\}\,,\quad \textrm{with}\,\xi^{LR}=0\,;
 \nonumber\\
A_{LR}: && \{\xi^{LL}=\xi^{LR},\phi_L\}=\{0.0104,-70^{\circ}\}\,,
 \qquad B_{LR}:  \{ \xi^{LL}= \xi^{LR},\phi_L\} =\{0.0186,83^{\circ}\}\,.
\end{eqnarray}
However, the corresponding prediction $A_{CP}(B^-_u\to\pi^0
K^-)=-0.03\pm0.01$~\cite{Barger} of solution $A_L$ and $A_{LR}$ in
Eq.~(\ref{Barger}) is obviously inconsistent with the up-to-date
experimental data $0.050\pm0.025$. Moreover, the annihilation
amplitudes, which could generate some strong-interaction phases, are
important for predicting CP violations.

Based on the above observations, in this paper we shall adopt the
QCDF approach and reevaluate the effects of the nonuniversal
$Z^{\prime}$ model on these decay modes with the updated
experimental data. Furthermore, since the $B\to\pi K^{\ast}$ and
$\rho K$ decays also involve the same quark level $b\to s \bar{q}q$
($q=u,d$) transition, it is necessary to take into account these
decay modes.

In Section~2, we provide a quick survey of $B\to\pi K$, $\pi
K^{\ast}$ and $\rho K$ decays in the SM within the QCDF formalism;
our numerical results, with two different schemes for the end-point
divergence, are also presented. In Section~3, after reviewing the
nonuniversal $Z^{\prime}$ model briefly, we present our analyses and
numerical results in detail. Section~4 contains our conclusions.
Appendix~A recapitulates the decay amplitudes for the twelve decay
modes within the SM~\cite{Beneke3}. Appendix~B contains the formulas
for hard-spectator and annihilation amplitudes with the infrared
finite gluon propagator~\cite{YDYang}. All  the theoretical input
parameters are summarized in Appendix~C.

\section{The SM results with two schemes for the end-point divergence.}
In the SM, the effective Hamiltonian responsible for $b\to s$
transitions is given as~\cite{Buchalla:1996vs}
\begin{eqnarray}\label{eq:eff}
 {\cal H}_{\rm eff} &=& \frac{G_F}{\sqrt{2}} \biggl[V_{ub}
 V_{us}^* \left(C_1 O_1^u + C_2 O_2^u \right) + V_{cb} V_{cs}^* \left(C_1
 O_1^c + C_2 O_2^c \right) - V_{tb} V_{ts}^*\, \big(\sum_{i = 3}^{10}
 C_i O_i \big. \biggl. \nonumber\\
 && \biggl. \big. + C_{7\gamma} O_{7\gamma} + C_{8g} O_{8g}\big)\biggl] +
 {\rm h.c.},
\end{eqnarray}
where $V_{qb} V_{qs}^*$~($q=u$, $c$ and $t$) are products of the
Cabibbo-Kobayashi-Maskawa~(CKM) matrix elements~\cite{ckm}, $C_{i}$
the Wilson coefficients, and $O_i$ the relevant four-quark operators
whose explicit forms could be found, for example, in
Refs.~\cite{Beneke2,Buchalla:1996vs}.

In recent years, the QCDF approach has been employed extensively to
study the hadronic B-meson decays. The $B\to\pi K$, $\pi K^{\ast}$
and $\rho K$ decays have been studied comprehensively within the SM
in Refs.~\cite{Beneke2, Beneke3, YD, DSDu}, and the relevant decay
amplitudes within this formalism are shown in Appendix A. It is also
noted that the framework contains estimates of the hard-spectator
and annihilation corrections. Even though they are power-suppressed,
their strength and associated strong-interaction phase are
numerically important to evaluate the branching ratio and the CP
asymmetry. However, unfortunately, the end-point singularities
appear in twist-3 spectator and annihilation amplitudes. So, how to
regulate the end-point divergence becomes important and necessary
within this formalism. Here we shall adopt the following two
schemes:

\subsubsection*{Scheme I: Parametrization}
As the most popular way, the end-point divergent integrals are
treated as signs of infrared sensitive contributions and
phenomenologically parameterized by~\cite{Beneke2,Beneke3}
\begin{equation}\label{treat-for-anni}
\int_0^1 \frac{\!dy}{y}\, \to X_A =(1+\rho_A e^{i\phi_A}) \ln
\frac{m_B}{\Lambda_h}, \qquad \int_0^1dy \frac{\textmd{ln}y}{y}\,
\to -\frac{1}{2}(X_A)^2\,,
 \end{equation}
with $\Lambda_h=0.5\,{\rm GeV}$, $\rho_A \leq 1$ and $\phi_A$
unrestricted. $X_{H}$ is treated in the same manner. The different
choices of $\rho_A$ and $\phi_A$ correspond to different scenarios
as discussed in Ref.~\cite{Beneke3}, and S4 is mentioned as the most
favorable one. It presents the moderate value of nonuniversal
annihilation phase $\phi_A=-55^{\circ}\,({\rm PP})$,
$-20^{\circ}\,({\rm PV})$ and $-70^{\circ}\,({\rm VP})$.
Conservatively, in our calculations we quote $\pm 5^{\circ}$ as
their theoretical uncertainties. Taking $\rho_A=1$ and $X_{A,H}$
universal for all decay processes belonging to the same modes~(PP,
PV or VP), we present our numerical results of branching ratios and
direct CP asymmetries for $ B\to\pi K$, $\pi K^{\ast}$ and $\rho K$
decays in the third column of Tables~\ref{tab_br} and \ref{tab_cp},
respectively.

As is known, the mixing-induced CP asymmetry $A^{mix}_{CP}$ is well
suited for testing the SM and searching for new physics effects. For
example, the investigation of mixing-induced CP asymmetries in
penguin dominated $\bar{B}^0\to\pi^0 K_S^{0}$ and
$\bar{B}^0\to\rho^0 K_S^{0}$ decay modes has attracted much
attention recently~\cite{benekeCP,Nir,Fleischer,Gronau}. After
neglecting the $K_0-\bar{K}_0$ mixing effect, the mixing-induced
asymmetry could be written as
\begin{equation}
 A^{mix}_{CP}(\bar{B}^0\to
 f)=\frac{2\,{\rm Im}\lambda_f}{1+|\lambda_f|^2}\,, \qquad
 (f=\pi^0 K_S^{0}\,, \rho^0 K_S^{0})\,,
\end{equation}
with $\lambda_f=-{\rm
exp}\{i\arg[\frac{V_{td}V_{tb}^{\ast}}{V_{td}^{\ast}V_{tb}}]\}\bar{A}_f/A_f$
in our phase convention. Our numerical predictions are listed in
Table~\ref{tab_mixcp}, which agree with the measurements within
large experimental errors.

\subsubsection*{Scheme II: Infrared finite dynamical gluon
propagator}
In our previous paper~\cite{YDYang}, we have thoroughly
studied the end-point divergence with an infrared finite dynamical
gluon propagator. It is interesting to note that recent theoretical
and phenomenological studies are now accumulating supports for a
softer infrared behavior of the gluon
propagator~\cite{Alkofer,theo,phe}. Furthermore, the infrared finite
dynamical gluon propagator, which is shown to be not divergent as
fast as $\frac{1}{q^2}$, has been successfully applied to the
hadronic B-meson decays~\cite{YYgluon, Natale}. In our evaluations,
we shall quote the gluon propagator derived by Cornwall (in
Minkowski space)~\cite{Cornwall}
\begin{eqnarray}
D(q^2)=\frac{1}{q^2-M_g^2(q^2)+i\epsilon}~,
 \label{Dg}
\end{eqnarray}
where $q$ is the gluon momentum. The corresponding strong coupling
constant reads
\begin{eqnarray}
\alpha_s(q^2)=\frac{4\pi}{\beta_0\mathrm{ln}\Big(\frac{q^2+4M_g^2(q^2)}
{\Lambda_{QCD}^2}\Big)}~, \label{Alphas}
\end{eqnarray}
where $\beta_0=11-\frac{2}{3}n_f$ is the first coefficient of the
beta function, with $n_f$ being the number of active quark flavors.
The dynamical gluon mass square $M_g^2(q^2)$ is obtained
as~\cite{Cornwall}
\begin{eqnarray}
M_g^2(q^2)=m_g^2\Bigg[\frac{\mathrm{ln}\Big(\frac{q^2+4m_g^2}{\Lambda_{QCD}^2}\Big)}
{\mathrm{ln}\Big(\frac{4m_g^2}{\Lambda_{QCD}^2}\Big)}\Bigg]^{-\frac{12}{11}},
\label{Mg}
\end{eqnarray}
where $m_g$ is the effective gluon mass and $\Lambda_{QCD}=225~{\rm
MeV}$. In Ref.~\cite{YDYang}, we present our suggestion,
$m_g=0.50\pm0.05~{\rm GeV}$, which is a reasonable choice so that
most of the observables (except for $A_{CP}(B\to\pi^0 K^-)$) are in
good agreement with the experimental data. In this way, we find that
the hard-spectator scattering contributions are real, and the
annihilation contributions are complex with a large imaginary
part~\cite{YDYang}. Our numerical predictions for branching ratios,
direct CP asymmetries and mixing-induced CP asymmetries are listed
in the fourth column of Tables~\ref{tab_br}, \ref{tab_cp} and
\ref{tab_mixcp}, respectively.

Although numerically these two schemes have some differences, both
of their predictions are consistent with most of the experimental
data within errors. However, as expected in the SM, we again find
that
$A_{CP}(B^{-}_u\to\pi^{0}K^{-})=-0.041\pm0.008~(-0.100\pm0.008)$,
are very close to $A_{CP}(\bar{B}^0_d\to\pi^{+}
K^{-})=-0.077\pm0.009~(-0.116\pm0.008)$ in the first~(second)
scheme. So, it is still hard to accommodate the measured large
difference $\Delta A$ in the SM within the QCDF formalism,
irrespective of adopting which scheme. In the following, we pursue
possible solutions to this problem with a family nonuniversal
$Z^{\prime}$ model~\cite{Langacker}.

\section{Solution to the ``$\pi K$ puzzle'' with nonuniversal $Z^{\prime}$ model.}

\subsection{Formalism of the family nonuniversal $Z^{\prime}$ model}
A possible heavy $Z^{\prime}$ boson is predicted in many extensions
of the SM, such as grand unified theories, superstring theories, and
theories with large extra dimensions. The simplest way to extend the
SM gauge structure is to include a new $U(1)$ gauge group. A family
nonuniversal $Z^{\prime}$ model can lead to FCNC processes even at
tree level due to the non-diagonal chiral coupling matrix. The
formalism of the model has been detailed in Ref.~\cite{Langacker}.
The relevant studies in the context of B physics have also been
extensively performed in Refs.~\cite{Barger,Barger1,BZprime,Giri}.

After neglecting the $Z-Z^{\prime}$ mixing with small mixing angle
$\theta\sim\mathcal {O}(10^{-3})$~\cite{Abreu}, and taking all the
fields being the physical eigenstates, the $Z^{\prime}$ part of the
neutral-current Lagrangian can be written as~\cite{Langacker}
\begin{equation}
\mathcal {L}^{\prime}=-g^{\prime}J_{\mu}^{\prime}Z^{\prime\mu}\,,
\end{equation}
where $g^{\prime}$ is the gauge coupling constant of extra
$U^{\prime}(1)$ group at the EW $M_W$ scale. The $Z^{\prime}$ chiral
current is
\begin{equation}
J_{\mu}^{\prime}=\bar{\psi}_i\gamma_{\mu}[(B_q^L)_{ij}P_L\,+\,(B_q^R)_{ij}P_R]\psi_j\,,
\end{equation}
where $\psi$ is the mass eigenstate of chiral fields and
$P_{L,R}=(1\mp\gamma_5)/2$. The effective chiral $Z^{\prime}$
coupling matrices are given as
\begin{equation}
 B_q^X=V_{qX}\epsilon_{qX}V_{qX}^{\dagger}\,,\qquad (q=u,d; X=L,R)\,.
\end{equation}

With the assumption of flavor-diagonal right-handed couplings , the
$Z^{\prime}$ part of the effective Hamiltonian for $b\to
s\bar{q}q\,(q=u,d)$ transitions can be written as~\cite{Barger}
\begin{equation}\label{heffz1}
 {\cal H}_{eff}^{\rm
 Z^{\prime}}=\frac{2G_F}{\sqrt{2}}\big(\frac{g^{\prime}M_Z}
 {g_1M_{Z^{\prime}}}\big)^2
 \,B_{sb}^L(\bar{s}b)_{V-A}\sum_{q}\big(B_{qq}^L (\bar{q}q)_{V-A}
 +B_{qq}^R(\bar{q}q)_{V+A}\big)+h.c.\,,
\end{equation}
where $g_1=e/(\sin{\theta_W}\cos{\theta_W})$ and $M_{Z^{\prime}}$
the new gauge boson mass. It is noted that the forms of the above
operators already exist in the SM. As a result, Eq.~(\ref{heffz1})
can be modified as
\begin{equation}
 {\cal H}_{eff}^{\rm
 Z^{\prime}}=-\frac{G_F}{\sqrt{2}}V_{ts}^{\ast}V_{tb}\sum_{q}
 (\Delta C_3 O_3^q +\Delta C_5 O_5^q+\Delta C_7 O_7^q+\Delta C_9
  O_9^q)+h.c.\,,
\end{equation}
where $O_i^q(i=3,5,7,9)$ are the effective operators in the SM, and
$\Delta C_i$ the modifications to the corresponding SM Wilson
coefficients caused by $Z^{\prime}$ boson, which are expressed as
\begin{eqnarray}
 \Delta C_{3,5}&=&-\frac{2}{3V_{ts}^{\ast}V_{tb}}\,\big(\frac{g^{\prime}M_Z}
 {g_1M_{Z^{\prime}}}\big)^2\,B_{sb}^L\,(B_{uu}^{L,R}+2B_{dd}^{L,R})\,,\nonumber\\
 \Delta C_{9,7}&=&-\frac{4}{3V_{ts}^{\ast}V_{tb}}\,\big(\frac{g^{\prime}M_Z}
 {g_1M_{Z^{\prime}}}\big)^2\,B_{sb}^L\,(B_{uu}^{L,R}-B_{dd}^{L,R})\,,
 \label{NPWilson}
\end{eqnarray}
in terms of the model parameters at the $M_W$ scale.

Generally, the diagonal elements of the effective coupling matrices
$B_{qq}^{L,R}$ are real as a result of the hermiticity of the
effective Hamiltonian. However, the off-diagonal ones of $B_{sb}^L$
can contain a new weak phase $\phi_L$. Then, conveniently we can
represent $\Delta C_i$ as\footnote{ For comparison,  
we take the same phase convention as Ref.~\cite{Barger}.}
\begin{eqnarray}
 \Delta
 C_{3,5}&=&2\,\frac{|V_{ts}^{\ast}V_{tb}|}{V_{ts}^{\ast}V_{tb}}\,
 \zeta^{LL,LR}\,e^{i\phi_L}\,,\nonumber\\
 \Delta
 C_{9,7}&=&4\,\frac{|V_{ts}^{\ast}V_{tb}|}{V_{ts}^{\ast}V_{tb}}\,
 \xi^{LL,LR}\,e^{i\phi_L}\,,
\end{eqnarray}
where the real NP parameters $\zeta^{LL,LR}$, $\xi^{LL,LR}$ and
$\phi_L$ are defined, respectively, as
\begin{eqnarray}
 \zeta^{LL,LR}&=&-\frac{1}{3}\,\big(\frac{g^{\prime}M_Z}
 {g_1M_{Z^{\prime}}}\big)^2\,\big|\frac{B_{sb}^L}{V_{ts}^{\ast}V_{tb}}\big|\,
 (B_{uu}^{L,R}+2B_{dd}^{L,R})\,,\nonumber\\
 \xi^{LL,LR}&=&-\frac{1}{3}\,\big(\frac{g^{\prime}M_Z}{g_1M_{Z^{\prime}}}\big)^2\,
 \big|\frac{B_{sb}^L}{V_{ts}^{\ast}V_{tb}}\big|\,(B_{uu}^{L,R}-B_{dd}^{L,R})\,,
 \nonumber\\
 \phi_L&=&{\rm Arg}[B_{sb}^L]\,.
\end{eqnarray}

\begin{table}[t]
 \begin{center}
 \caption{The Wilson coefficients $C_i$ within the SM and with the contribution
 from $Z^{\prime}$ boson included in NDR scheme at the scale $\mu=m_{b}$ and
 $\mu_h=\sqrt{\Lambda_h m_b}$.}
 \label{Wilson}
 \vspace{0.1cm}
 \small
 \doublerulesep 0.7pt \tabcolsep 0.04in
 \begin{tabular}{c|cc|cc}\hline\hline
 Wilson                   &\multicolumn{2}{c|}{$\mu=m_{b}$}                        &\multicolumn{2}{c}{$\mu_h=\sqrt{\Lambda_h m_b}$} \\\cline{2-3}\cline{4-5}
 coefficients             &$C_i^{SM}$ &$\Delta C_i^{Z^{\prime}}$                   &$C_i^{SM}$ &$\Delta C_i^{Z^{\prime}}$\\ \hline\hline
 $C_1$                    &$1.075$    &$-0.006X$                                   &$1.166$    &$-0.008X$\\
 $C_2$                    &$-0.170$   &$-0.009X$                                   &$-0.336$   &$-0.014X$ \\
 $C_3$                    &$0.013$    &$0.05X-0.01Y-2.20X^{\prime}-0.05Y^{\prime}$ &$0.025$    &$0.11X-0.02Y-2.37X^{\prime}-0.12Y^{\prime}$\\
 $C_4$                    &$-0.033$   &$-0.13X+0.01Y+0.55X^{\prime}+0.02Y^{\prime}$&$-0.057$   &$-0.24X+0.02Y+0.92X^{\prime}+0.09Y^{\prime}$\\
 $C_5$                    &$0.008$    &$0.03X+0.01Y-0.06X^{\prime}-1.83Y^{\prime}$ &$0.011$    &$0.03X+0.02Y-0.10X^{\prime}+0.09Y^{\prime}$\\
 $C_6$                    &$-0.038$   &$-0.15X+0.01Y+0.1X^{\prime}-0.6Y^{\prime}$  &$-0.076$   &$-0.32X+0.04Y+0.16X^{\prime}-1.26Y^{\prime}$\\
 $C_7/{\alpha}_{em}$      &$-0.015$   &$4.18X-473Y+0.25X^{\prime}+1.27Y^{\prime}$  &$-0.034$   &$5.7X-459Y+0.4X^{\prime}+1.7Y^{\prime}$\\
 $C_8/{\alpha}_{em}$      &$0.045$    &$1.18X-166Y+0.01X^{\prime}+0.56Y^{\prime}$  &$0.089$    &$3.2X-355Y+0.2X^{\prime}+1.5Y^{\prime}$\\
 $C_9/{\alpha}_{em}$      &$-1.119$   &$-561X+4.52Y-0.8X^{\prime}+0.4Y^{\prime}$   &$-1.228$   &$-611X+6.7Y-1.2X^{\prime}+0.6Y^{\prime}$\\
 $C_{10}/{\alpha}_{em}$   &$0.190$    &$118X-0.5Y+0.2X^{\prime}-0.05Y^{\prime}$    &$0.356$    &$207X-1.4Y+0.5X^{\prime}-0.1Y^{\prime}$\\
 $C_{7{\gamma}}$          &$-0.297$   &---                                         &$0.360$    &---\\
 $C_{8g}$                 &$-0.143$   &---                                         &$-0.168$   &---\\
 \hline \hline
 \end{tabular}
 \end{center}
 \end{table}

It is noted that the other SM Wilson coefficients may also receive
contributions from the $Z^{\prime}$ boson through renormalization
group~(RG) evolution. With our assumption that no significant RG
running effect between $M_Z^{\prime}$ and $M_W$ scales, the RG
evolution of the modified Wilson coefficients is exactly the same as
the ones in the SM~\cite{Buchalla:1996vs,Buras:2000}. For
simplicity, we define
\begin{eqnarray}
 X^{\prime}=\zeta^{LL}e^{i\phi_L}\,,\qquad
 Y^{\prime}=\zeta^{LR}e^{i\phi_L}\,, \nonumber\\
 X=\xi^{LL}e^{i\phi_L}\,,\qquad Y=\xi^{LR}e^{i\phi_L}\,.
\end{eqnarray}
The numerical results of Wilson coefficients in the naive
dimensional regularization~(NDR) scheme at the scale
$\mu=m_{b}$~($\mu_h=\sqrt{\Lambda_h m_b}$) are listed in
Table~\ref{Wilson}. The values at the scale $\mu_{h}$, with
$m_{b}=4.79~{\rm GeV}$ and $\Lambda_{h}=500~{\rm MeV}$, should be
used in the calculation of hard-spectator and weak annihilation
contributions.

\begin{table}[t]
 \begin{center}
 \caption{The $CP$-averaged branching ratios~(in units of $10^{-6}$)
 of $B$ ${\to}$ $\pi K$, $\pi K^{\ast}$ and $\rho K$ decays in the SM
 with two end-point divergence regulation schemes, and in the nonuniversal
 $Z^{\prime}$ model with four different cases.}
 \label{tab_br}
 \vspace{0.5cm}
 \small
 \doublerulesep 0.7pt \tabcolsep 0.05in
 \begin{tabular}{lcccccccccccc} \hline \hline
 \multicolumn{1}{c}{Decay Mode}             &\multicolumn{1}{c}{Exp.}&\multicolumn{2}{c}{ SM }&\multicolumn{4}{c}{$Z^{\prime}$ model} \\
                                                                      &     data     & Scheme I     & Scheme II   &Case I       &Case II      &Case III     &Case IV \\ \hline
 $B_{u}^{-}$ ${\to}$ ${\pi}^{-}{\overline{K}}^{0}$                     &$23.1\pm1.0$ &$19.0\pm2.5$  &$23.4\pm3.9$ &$23.3\pm0.7$ &$23.3\pm0.6$ &$23.2\pm0.6$ &$23.3\pm0.7$\\
 $B_{u}^{-}$ ${\to}$ ${\pi}^{0}K^{-}$                                  &$12.9\pm0.6$ &$10.5\pm1.3$  &$12.7\pm2.0$ &$12.5\pm0.6$ &$12.6\pm0.6$ &$12.5\pm0.5$ &$12.6\pm0.6$\\
 ${\overline{B}}_{d}^{0}$ ${\to}$ ${\pi}^{+}K^{-}$                     &$19.4\pm0.6$ &$16.2\pm2.2$  &$20.1\pm3.4$ &$19.9\pm0.5$ &$19.8\pm0.5$ &$19.9\pm0.5$ &$20.0\pm0.5$\\
 ${\overline{B}}_{d}^{0}$ ${\to}$ ${\pi}^{0}{\overline{K}}^{0}$        &$9.8\pm0.6$  &$7.3\pm1.1$   &$9.3\pm1.7$  &$9.4\pm0.6$  &$9.5\pm0.6$  &$9.1\pm0.4$  &$9.1\pm0.4$\\
  \hline
 $B_{u}^{-}$ ${\to}$ ${\pi}^{-}{\overline{K}}^{{\ast}0}$               &$10.0\pm0.8$ &$11.7\pm1.2$  &$10.3\pm3.3$ &$8.4\pm1.0$  &$8.5\pm0.9$  &$8.7\pm0.6$  &$8.6\pm0.7$\\
 $B_{u}^{-}$ ${\to}$ ${\pi}^{0}K^{{\ast}-}$                            &$6.9\pm2.3$  &$7.0\pm0.7$   &$6.0\pm1.8$  &$4.7\pm0.6$  &$4.7\pm0.5$  &$4.9\pm0.3$  &$4.8\pm0.3$\\
 ${\overline{B}}_{d}^{0}$ ${\to}$ ${\pi}^{+}K^{{\ast}-}$               &$10.3\pm 1.1$&$9.9\pm1.1$   &$9.2\pm2.8$  &$7.5\pm1.0$  &$7.7\pm0.9$  &$8.0\pm0.6$  &$8.0\pm0.6$\\
 ${\overline{B}}_{d}^{0}$ ${\to}$ ${\pi}^{0}{\overline{K}}^{{\ast}0}$  &$2.4\pm0.7$  &$4.1\pm0.5$   &$3.9\pm1.3$  &$3.6\pm0.5$  &$3.7\pm0.4$  &$3.5\pm0.4$  &$3.5\pm0.4$\\
  \hline

 $B_{u}^{-}$ ${\to}$ ${\rho}^{-}{\overline{K}}^{0}$              &$8.0^{+1.5}_{-1.4}$&$5.2\pm0.9$   &$10.6\pm3.7$ &$9.6\pm1.4$  &$9.7\pm1.3$   &$10.6\pm1.3$&$10.7\pm1.5$\\
 $B_{u}^{-}$ ${\to}$ ${\rho}^{0}K^{-}$                        &$3.81^{+0.48}_{-0.46}$&$2.5\pm0.4$   &$5.4\pm1.6$  &$4.22\pm0.62$ &$4.47\pm0.63$&$4.7\pm0.6$ &$4.8\pm0.7$\\
 ${\overline{B}}_{d}^{0}$ ${\to}$ ${\rho}^{+}K^{-}$             &$8.6^{+0.9}_{-1.1 }$&$6.3\pm1.0$   &$13.0\pm3.8$ &$10.8\pm1.4$ &$10.9\pm1.4$  &$11.9\pm1.4$&$12.5\pm1.8$\\
 ${\overline{B}}_{d}^{0}$ ${\to}$ ${\rho}^{0}{\overline{K}}^{0}$&$5.4^{+0.9}_{-1.0 }$&$3.7\pm0.5$   &$7.3\pm2.1$  &$7.1\pm0.9$  &$7.4\pm0.9$   &$6.8\pm0.9$ &$6.9\pm1.0$\\
 \hline \hline
 \end{tabular}
 \end{center}
 \end{table}

\begin{table}[t]
 \begin{center}
 \caption{The direct CP asymmetries~(in unit of $10^{-2}$) of $B$
 ${\to}$ $\pi K$, $\pi K^{\ast}$ and $\rho K$ decays. The other captions
 are the same as Table.~\ref{tab_br}.}
 \label{tab_cp}
 \vspace{0.1cm}
 \small
 \doublerulesep 0.7pt \tabcolsep 0.04in
 \begin{tabular}{lccccccc} \hline \hline
 \multicolumn{1}{c}{Decay Mode}&\multicolumn{1}{c}{Exp.}&\multicolumn{2}{c}{SM} &\multicolumn{4}{c}{ $Z^{\prime}$ model}\\
                                                                      & data       & Scheme I    & Scheme II    &Case I        &Case II      &Case III     &Case IV\\ \hline
 $B_{u}^{-}$ ${\to}$ ${\pi}^{-}{\overline{K}}^{0}$                    &$0.9\pm2.5$ &$0.4\pm0.1$  &$0.04\pm0.07$ &$-1.6\pm0.3$  &$-2.7\pm0.9$ &$5.2\pm0.5$  &$5.1\pm0.6$\\
 $B_{u}^{-}$ ${\to}$ ${\pi}^{0}K^{-}$                                 &$5.0\pm2.5$ &$-4.1\pm0.8$ &$-10.0\pm0.8$ &$2.4\pm1.6$   &$2.3\pm1.5$  &$0.9\pm0.7$  &$1.2\pm0.9$\\
 ${\overline{B}}_{d}^{0}$ ${\to}$ ${\pi}^{+}K^{-}$           &$-9.8^{+1.2}_{-1.1}$ &$-7.7\pm0.9$ &$-11.6\pm0.3$ &$-11.7\pm0.3$ &$-11.0\pm0.7$&$-10.5\pm1.1$&$-10.5\pm1.2$\\
 ${\overline{B}}_{d}^{0}$ ${\to}$ ${\pi}^{0}{\overline{K}}^{0}$       &$-1\pm10$   &$-1.5\pm0.3$ &$0.7\pm0.3$   &$-17\pm2$     &$-18\pm2$    &$-6\pm2$     &$-6\pm2$\\
 \hline
 $B_{u}^{-}$ ${\to}$ ${\pi}^{-}{\overline{K}}^{{\ast}0}$       &$-2^{+6.7}_{-6.1}$ &$0.6\pm0.1$  &$0.09\pm0.15$ &$-2.1\pm0.4$  &$-3.3\pm0.5$ &$-0.6\pm2.4$ &$-3.0\pm6.7$\\
 $B_{u}^{-}$ ${\to}$ ${\pi}^{0}K^{{\ast}-}$                           &$4\pm29$    &$-6\pm2$     &$-37\pm9$     &$6.8\pm7.1$   &$9.1\pm7.2$  &$-17\pm4$    &$-18\pm6$\\
 ${\overline{B}}_{d}^{0}$ ${\to}$ ${\pi}^{+}K^{{\ast}-}$              &$-25\pm11$  &$-13\pm2$    &$-43\pm10$    &$-48\pm3$     &$-46\pm3$    &$-49\pm3$    &$-50\pm5$\\
 ${\overline{B}}_{d}^{0}$ ${\to}$ ${\pi}^{0}{\overline{K}}^{{\ast}0}$ &$-15\pm12$  &$-4\pm1$     &$4\pm2$       &$-58\pm9$     &$-62\pm9$    &$-34\pm7$    &$-36\pm11$\\
  \hline
 $B_{u}^{-}$ ${\to}$ ${\rho}^{-}{\overline{K}}^{0}$                   &$-12\pm17$  &$0.4\pm0.1$  &$0.5\pm0.2$   &$1.5\pm0.1$   &$-0.15\pm0.7$&$5.3\pm1.1$  &$6.5\pm4.5$\\
 $B_{u}^{-}$ ${\to}$ ${\rho}^{0}K^{-}$                      &$41.9^{+8.1}_{-10.4}$ &$57.3\pm5.8$ &$42.3\pm9.5$  &$-36\pm10$    &$-46\pm12$   &$27\pm4$     &$27\pm5$\\
 ${\overline{B}}_{d}^{0}$ ${\to}$ ${\rho}^{+}K^{-}$                   &$15\pm6$    &$36\pm4$     &$29\pm6$      &$31\pm3$      &$33\pm3$     &$25\pm2$     &$25\pm2$\\
 ${\overline{B}}_{d}^{0}$ ${\to}$ ${\rho}^{0}{\overline{K}}^{0}$      &$1\pm20$   &$-2.1\pm1.3$ &$-2.4\pm1.4$  &$45\pm5$      &$50\pm5$     &$8\pm3$      &$9\pm4$\\
 \hline \hline
 \end{tabular}
 \end{center}
 \end{table}

\begin{table}[t]
 \begin{center}
 \caption{The mixing-induced CP asymmetries~(in unit of $10^{-2}$)
 of $\bar{B}^0\to\pi^0K_S^{0}$ and $\rho^0K_S^{0}$ decays. The other captions
 are the same as Table.~\ref{tab_br}.}
 \label{tab_mixcp}
 \vspace{0.5cm}
 \doublerulesep 0.7pt \tabcolsep 0.07in
 \begin{tabular}{lccccccc} \hline \hline
 \multicolumn{1}{c}{Decay Mode}     &\multicolumn{1}{c}{Exp.}    &\multicolumn{2}{c}{SM} &\multicolumn{4}{c}{$Z^{\prime}$ model}\\
                                                     & data            &Scheme I  &Scheme II &Case I   &Case II &Case III&Case IV\\ \hline
 ${\overline{B}}_{d}^{0}$ ${\to}$ ${\pi}^{0}K_S^{0}$ &$57\pm17$        &$77\pm2$  &$77\pm2$  &$46\pm6$ &$44\pm6$&$61\pm3$&$62\pm5$\\
 ${\overline{B}}_{d}^{0}$ ${\to}$ ${\rho}^{0}K_S^{0}$&$63^{+17}_{-21}$ &$60\pm2$  &$66\pm2$  &$87\pm2$ &$84\pm3$&$85\pm3$&$86\pm9$\\
 \hline \hline
 \end{tabular}
 \end{center}
 \end{table}

\subsection{Numerical analyses and discussions}

\begin{figure}[t]
\begin{center}
\epsfxsize=15cm \centerline{\epsffile{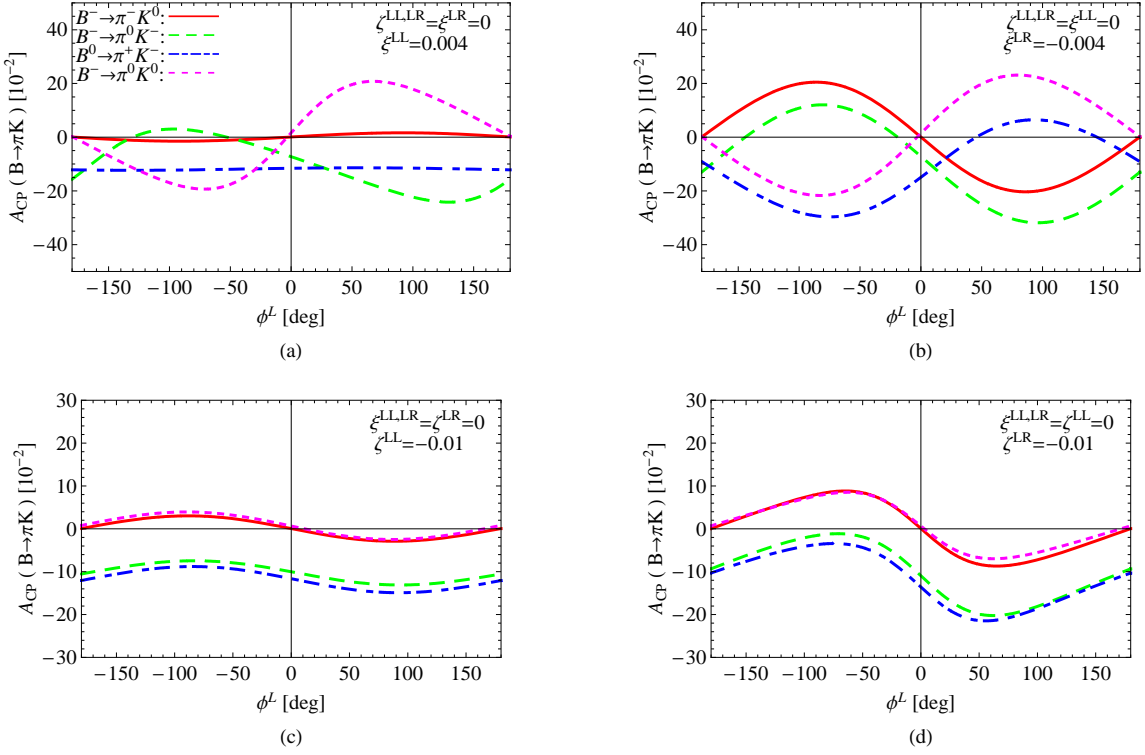}}
\centerline{\parbox{16cm}{\caption{\label{Acppik}\small The
dependence of $A_{CP}(B\to\pi K)$ on the new weak phase $\phi_L$
for  the values of $\xi^{LL}$, $\xi^{LR}$, $\zeta^{LL}$, and
$\zeta^{LR}$ as marked by  the legends. }}}
\end{center}
\end{figure}
\begin{figure}[t]
\begin{center}
\epsfxsize=15cm \centerline{\epsffile{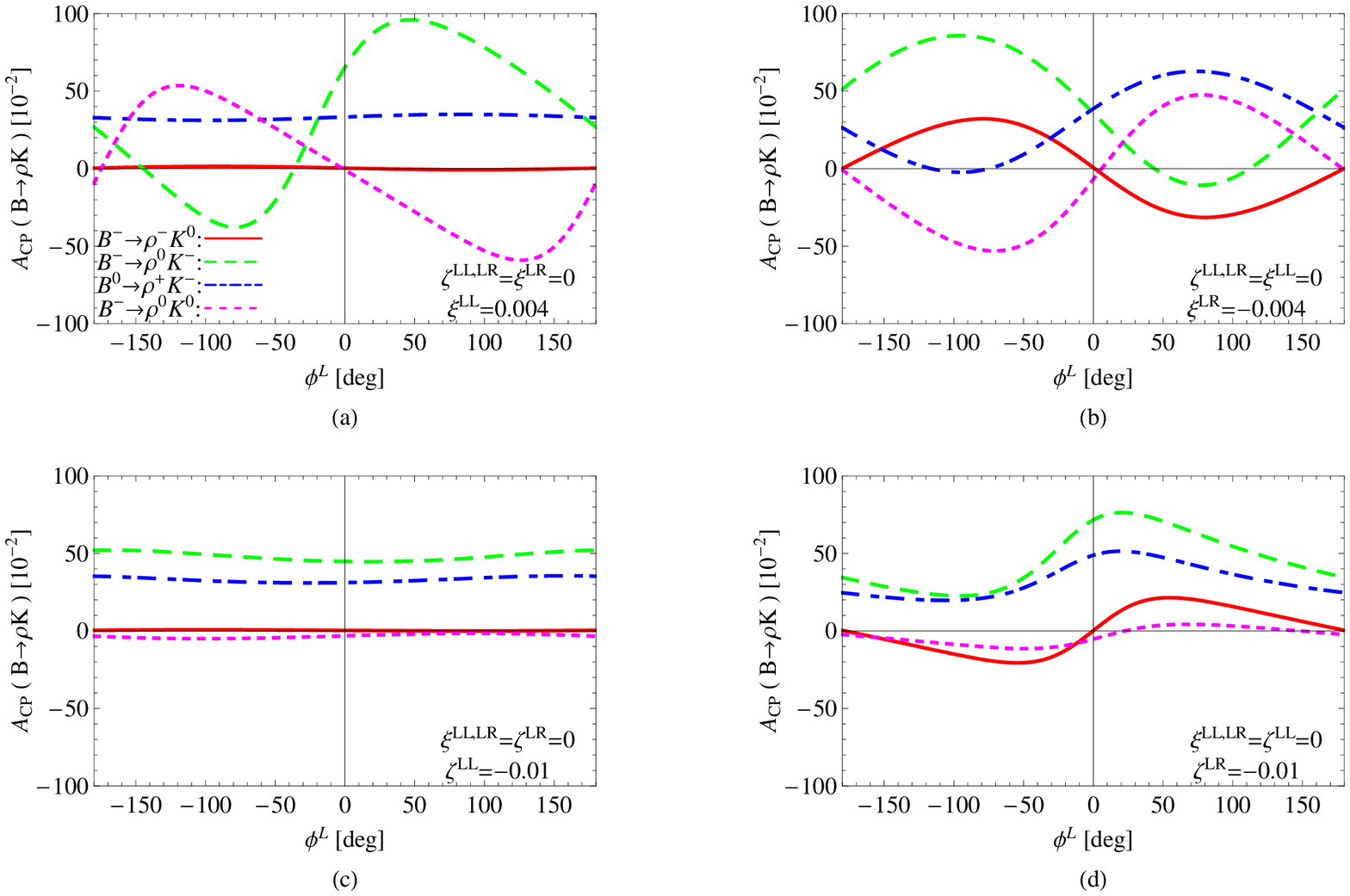}}
\centerline{\parbox{16cm}{\caption{\label{Acprhok}\small The
dependence of $A_{CP}(B\to\rho K)$ on the new weak phase $\phi_L$.}}}
\end{center}
\end{figure}

With the theoretical formulas and the input parameters summarized in
Appendix A, B and C, we now present our numerical analyses and
discussions. Our analyses are divided into the following four cases
with different simplifications for our attention, namely,
\begin{center}
\begin{itemize}
\item Case I:   With the simplifications $B_{uu}^{L,R}\simeq-2B_{dd}^{L,R}$
({\it i.e., $\zeta^{LL,LR}=0$}) and $\xi^{LR}=0$,

\item Case II:  With the simplifications $B_{uu}^{L,R}\simeq-2B_{dd}^{L,R}$
only ({\it i.e.,} $\zeta^{LL,LR}=0$),

\item Case III: Taking $B_{uu}^{R}\simeq-2B_{dd}^{R}$~({\it i.e.,} $\zeta^{LR}
\simeq0$), and leaving $\zeta^{LL}$ and $\xi^{LL,LR}$ arbitrary,

\item Case IV:  Without any simplifications for $B_{uu}^{L,R}$ and
$B_{dd}^{L,R}$, {\it i.e.,} arbitrary values for $\zeta^{LL,LR}$ and
$\xi^{LL,LR}$ are allowed.
\end{itemize}
\end{center}

Our fitting is performed with the experimental data varying randomly
within their $2\sigma$ error-bars, while the theoretical
uncertainties are obtained by varying the input parameters within
the regions specified in Appendix C. In addition, we quote the
Scheme II (taking $m_g=0.5{\rm GeV}$) to regulate the appearing
end-point divergences.

With the assumption $B_{uu}^{L,R}\simeq-2B_{dd}^{L,R}$ and
neglecting the color-suppressed EW penguins and the annihilation
amplitudes, four possible solutions Eq.~(\ref{Barger}) to the ``$\pi
K$ puzzle'' are obtained in Ref.~\cite{Barger}. It is still worth to
recheck these solutions with the updated experiment data and taken
into account the neglected corrections. Furthermore, the possible
solutions may also suffer strong constraints from $B\to\pi K^{\ast}$
and $\rho K$ decays, since they are also mediated by the same quark
level $b\to s\bar{q}q$ transitions.

\subsubsection*{Case I: With the simplifications
$B_{uu}^{L,R}\simeq-2B_{dd}^{L,R}$({\it i.e., $\zeta^{LL,LR}=0$})
and $\xi^{LR}=0$}

\begin{figure}[t]
\begin{center}
\epsfxsize=8cm \centerline{\epsffile{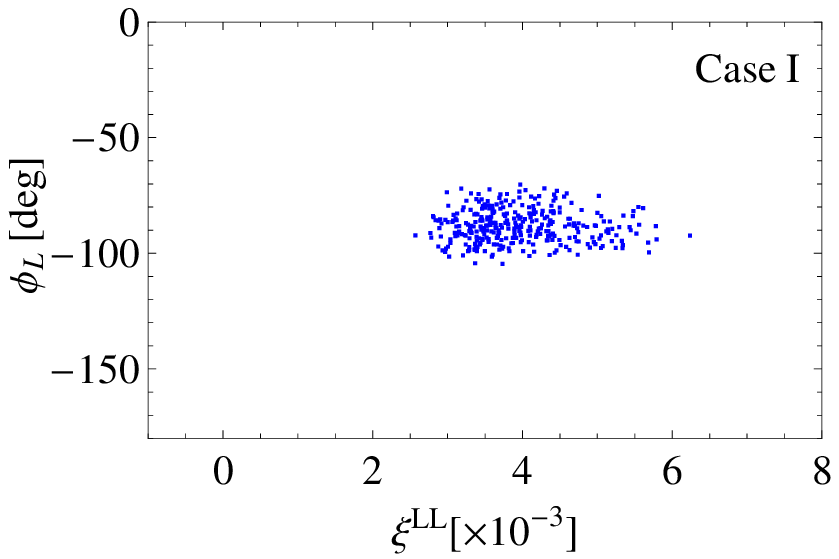}}
\centerline{\parbox{16cm}{\caption{\label{CaseI}\small The allowed
regions for the parameters $\xi^{LL}$ and $\phi_L$ in Case I.}}}
\end{center}
\end{figure}

In this case, assuming $B_{uu}^{L,R}\simeq-2B_{dd}^{L,R}$ as in
Ref.~\cite{Barger}, the NP effect primarily manifests itself in the
EW penguin sector and the $Z^{\prime}$ contribution to the Wilson
coefficients Eq.~(\ref{NPWilson}) can be simplified as
\begin{eqnarray}
 \Delta C_{3,5}&=&0\,,\nonumber\\
 \Delta
 C_{9,7}&=&4\,\frac{|V_{ts}^{\ast}V_{tb}|}{V_{ts}^{\ast}V_{tb}}\,
 \xi^{LL,LR}e^{i\phi_L}\,,\quad \textrm{with}\,\xi^{LL,LR}=\big(
 \frac{g^{\prime}M_Z}{g_1M_{Z^{\prime}}}\big)^2\,\big|\frac{B_{sb}^L}
 {V_{ts}^{\ast}V_{tb}}\big|\,B_{dd}^{L,R}\,.
\end{eqnarray}

As shown in Fig.~\ref{Acppik}~(a), taking $\xi^{LL}=0.004$ and
$\xi^{LR}=0$, we find that $A_{CP}(B^-\to\pi^0 K^-)$ is enhanced to
be consistent with the experimental data when
$\phi_L\sim-90^{\circ}$. Moreover, $A_{CP}(B^-\to\pi^- K^0)$ and
$A_{CP}(B^0\to\pi^+ K^-)$, which agree roughly with the experimental
data in the SM, are not sensitive to the parameter $\xi^{LL}$. So, a
possible solution to the observed ``$\pi K$ puzzle''
Eq.~(\ref{cpdiff}) in Case I is naively favored.

Taking ${\cal B}(B\to\pi K)$ and $A_{CP}(B\to\pi K)$ as constraints
on $\xi^{LL}$ and $\phi_L$, the allowed region for these two
parameters are shown in Fig.~\ref{CaseI} and the corresponding
numerical results are listed in Table.~\ref{NPPareValue}, {\it
i.e.,} $\xi^{LL}=(3.96\pm0.70)\times10^{-3}$ and
$\phi_L=-88^{\circ}\pm7^{\circ}$. Our result confirms that the
solution $B_L$ in Eq.~(\ref{Barger}) is helpful to resolve the
``$\pi K$ puzzle''~(note that a bit of difference might be due to
the fact that the annihilation corrections are not included in
Ref.~\cite{Barger}). However, the solution $A_L$ is excluded by the
updated experimental data
$A_{CP}(B^{-}\to\pi^{0}K^{-})=0.050\pm0.025$ as indicated in
Fig.~\ref{Acppik}~(a).

With $\xi^{LL}=(3.96\pm0.70)\times10^{-3}$ and
$\phi_L=-88^{\circ}\pm7^{\circ}$ as input parameters, we present our
predictions for ${\cal B}(B\to\pi K^{\ast}, \rho K)$,
$A_{CP}(B\to\pi K^{\ast}, \rho K)$ and $A_{CP}^{mix}(B^0\to\pi^0
K_S, \rho^0 K_S)$ in the fifth column of Tables.~\ref{tab_br},
\ref{tab_cp} and \ref{tab_mixcp}, respectively. We can see that most
of them are consistent with the experimental data within $2\sigma$.
Especially, the predicted $A_{CP}^{mix}(B^0\to\pi^0
K_S)=0.46\pm0.06$ is very close to the measurement
$0.57\pm0.17$~\cite{HFAG}. However, the prediction for
$A_{CP}(B^-\to\rho^0 K^-)=-0.36\pm0.10$ presents a large
discrepancy~(larger than $6\sigma$ errors) with the current
experiment data $0.419^{+0.081}_{-0.104}$~\cite{HFAG}, which is also
shown in Fig.~\ref{Acprhok}~(a). This fact implies that
$A_{CP}(B^-\to\rho^0 K^-)$ can provide a strong constraint on the
$Z^{\prime}$ couplings, at lease in Case I, and some more general
$Z^{\prime}$ models might be required to explain all of these
measurements.

\subsubsection*{Case II: With the simplification
$B_{uu}^{L,R}\simeq-2B_{dd}^{L,R}$ only ({\it i.~e.}
$\zeta^{LL,LR}=0$).}

\begin{figure}[t]
\begin{center}
\epsfxsize=15cm \centerline{\epsffile{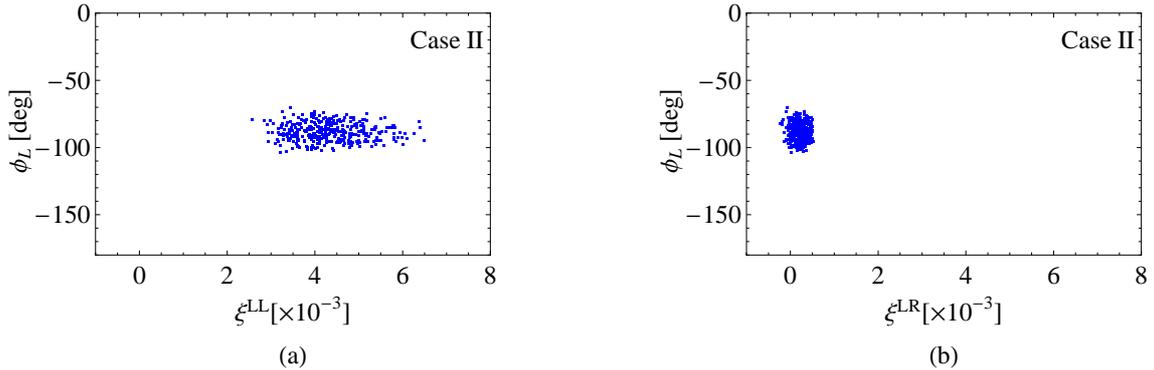}}
\centerline{\parbox{15cm}{\caption{\label{CaseII}\small The allowed
regions for the parameters $\xi^{LL,LR}$ and $\phi_L$ in Case II.}}}
\end{center}
\end{figure}

It is interesting to note that, as shown in Fig.~\ref{Acppik}~(b), a
region of minus $\xi^{LR}$ with $\phi_L\sim-90^{\circ}$ can bridge
the discrepancy of $A_{CP}(B^-\to\pi^0 K^-)$ between theoretical
predictions and experimental data. Moreover, it is also possible to
moderate the problem of $A_{CP}(B^-\to\rho^0 K^-)$ induced by
$\xi^{LL}$ as shown in Fig.~\ref{Acprhok}~(b). So, in Case II we
give up the simplification $\xi^{LR}=0$ and pursue possible
solutions to these discrepancies.

Taking ${\cal B}(B\to\pi K)$ and $A_{CP}(B\to\pi K)$ as constraints,
we present the allowed regions for $\xi^{LL}$, $\xi^{LR}$ and
$\phi_L$ in Fig.~\ref{CaseII}. Unfortunately, we find that the
required region of minus $\xi^{LR}$ with $\phi_L\sim-90^{\circ}$ is
excluded by $A_{CP}(B^{0}\to\pi^{+}K^{-})$, because it will induce a
large negative $A_{CP}(B^{0}\to\pi^{+}K^{-})$ as shown in
Fig.~\ref{Acppik}~(b). In addition, as shown in
Fig.~\ref{Acppik}~(b), the region of plus $\xi^{LR}$ with
$\phi_L\sim-90^{\circ}$ is helpless to resolve the ``$\pi K$
puzzle''. The $Z^{\prime}$ effects are therefore still dominated by
large $\xi^{LL}$, and the problem of $A_{CP}(B^-\to\rho^0 K^-)$
induced by $\xi^{LL}$ still exist.

In fact, with $\xi^{LL}$ and $\xi^{LR}$ having the same sign, the
corresponding $Z^{\prime}$ contributions counteract with each other
in the $B\to \pi^0 K^-$ decay as shown in Figs.~\ref{Acppik}~(a) and
(b). It is also easily understood from the expression for the
effective coefficient
$\alpha_{3,EW}^p(PP)=a_9^p(PP)-a_7^p(PP)$~\cite{Beneke3}, which
involves the leading-order $Z^{\prime}$ contribution in this case.
Thus, we conclude that any attempt to explain the $B\to\pi K$
anomaly in the non-universal $Z^{\prime}$ model with the assumption
$\xi^{LL}=\xi^{LR}=\xi$, as made in Ref.~\cite{Giri}, is frangible
and excluded in our case.

In a word, although the $Z^{\prime}$ contributions with a positive
$\xi^{LL}$ or a negative $\xi^{LR}$ and $\phi_L\sim-90^{\circ}$ are
helpful to bridge the discrepancy of $A_{CP}(B^-\to\pi^0 K^-)$, they
would induce the unmatched $A_{CP}(B^-\to\rho^0 K^-)$ and
$A_{CP}(B^0\to\pi^+ K^-)$, respectively. Thus, with both ${\cal
B}(B\to\pi K)$ and $A_{CP}(B\to\pi K, \rho K)$ as constraints, our
results indicate that all of the parameter spaces in Case I and Case
II are excluded with the assumption
$B_{uu}^{L,R}\simeq-2B_{dd}^{L,R}$. As an alternative, in the
following, we proceed to pursue possible solutions to these
observations by considering the $Z^\prime$ contributions to the QCD
penguins $\triangle C_{3,5}$.

\begin{table}[t]
 \begin{center}
 \caption{The numerical results for the parameters $\xi^{LL,LR}$,
 $\zeta^{LL,LR}$ and $\phi_L$ in the four different cases. The dashes
 mean that the corresponding parameters are neglected in each case.}
 \label{NPPareValue}
 \vspace{0.5cm}
 \doublerulesep 0.7pt \tabcolsep 0.07in
 \begin{tabular}{lccccccccccc} \hline \hline
  Parameters                  &Case I                   &Case II                  &Case III                   &Case IV\\\hline
 $\xi^{LL}(\times 10^{-3})$   &$3.96\pm0.70$            &$4.32\pm0.75$            &$1.52\pm0.24$              &$1.65\pm0.35$\\
 $\xi^{LR}(\times 10^{-3})$   &---                      &$0.21\pm0.15$            &$-0.53\pm0.13$             &$-0.54\pm0.15$ \\
 $\zeta^{LL}(\times 10^{-3})$ &---                      &---                      &$-11.8\pm3.1$              &$-14.6\pm7.1$\\
 $\zeta^{LR}(\times 10^{-3})$ &---                      &---                      &---                        &$1.04\pm2.70$\\
 $\phi^{L}$                   &$-88^{\circ}\pm7^{\circ}$&$-88^{\circ}\pm7^{\circ}$&$-86^{\circ}\pm14^{\circ}$ &$-85^{\circ}\pm16^{\circ}$\\
 \hline \hline
 \end{tabular}
 \end{center}
 \end{table}

\begin{figure}[t]
\begin{center}
\epsfxsize=15cm \centerline{\epsffile{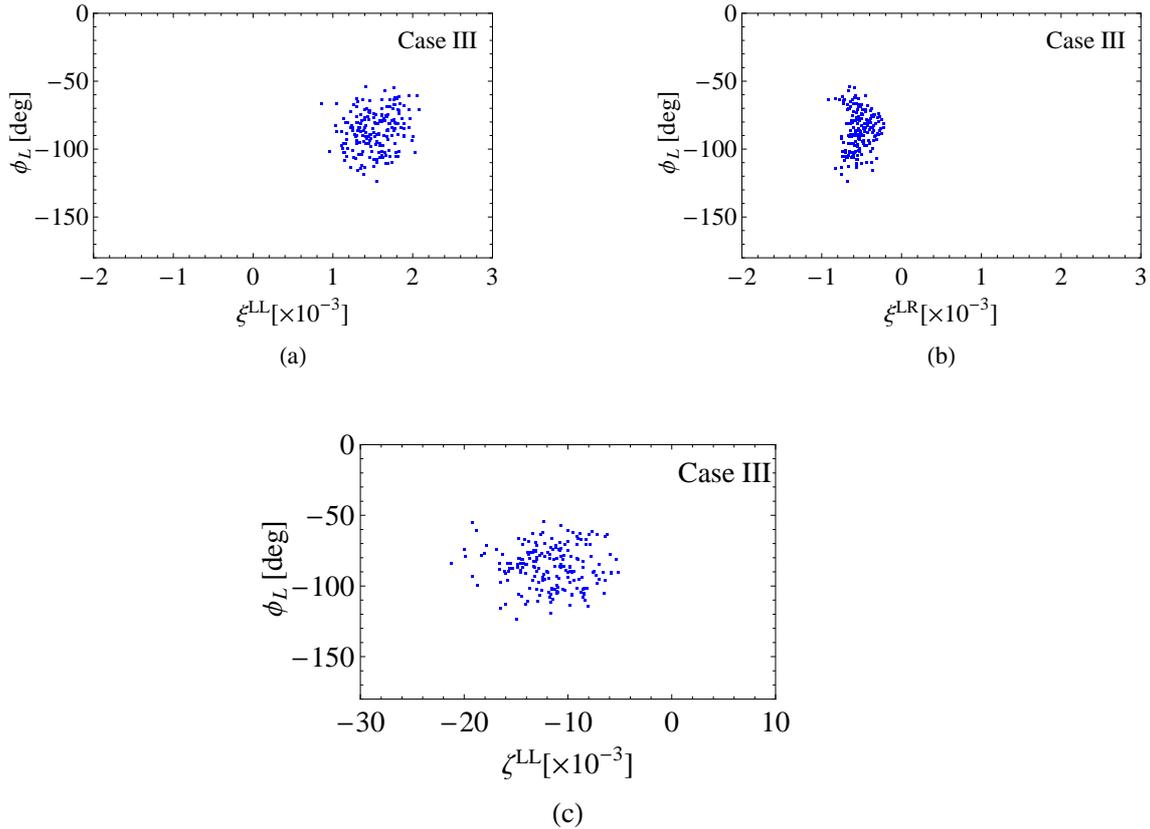}}
\centerline{\parbox{16cm}{\caption{\label{CaseIII}\small The allowed
regions for the parameters $\xi^{LL,LR}$, $\zeta^{LL}$, and $\phi_L$
in Case III.}}}
\end{center}
\end{figure}

\subsubsection*{Case III: Taking $B_{uu}^{R}\simeq-2B_{dd}^{R}$({\it i.e.,}
$\zeta^{LR}\simeq0$), and leaving $\zeta^{LL}$ and $\xi^{LL,LR}$
arbitrary.}

As shown in Fig.~\ref{Acppik}~(c), we find that the variation trends
of $A_{CP}(B^0\to\pi^+ K^-)$ and $A_{CP}(B^-\to\pi^0 K^-)$ are
always the same, indicating that the $Z^{\prime}$ contributions in
this case could not give a solution to the observed ``$\pi K$
puzzle'' directly, as well as the unmatched $A_{CP}(B^-\to\rho^0
K^-)$ induced by $\xi^{LL}$. However, it is interesting to note
that, with $\phi_L\sim-90^{\circ}$, both $A_{CP}(B^0\to\pi^+ K^-)$
and $A_{CP}(B^-\to\pi^0 K^-)$ could be enhanced simultaneously,
which may relax the constraints on $\xi^{LR}$. As mentioned in Case
II, a negative $\xi^{LR}$ is favored by the ``$\pi K$ puzzle'' and
can moderate the problem of $A_{CP}(B^-\to\rho^0 K^-)$ induced by
$\xi^{LL}$. So, the parameter $\zeta^{LL}$ may play an important
role.

With ${\cal B}(B\to\pi K)$, $A_{CP}(B\to\pi K)$ and $A_{CP}(B\to\rho
K)$ as constraints, the allowed regions for $\xi^{LL}$, $\xi^{LR}$,
$\zeta^{LL}$ and $\phi_L$ are shown in Figs.~\ref{CaseIII}. We find
that none of $\xi^{LL}$, $\xi^{LR}$ and $\zeta^{LL}$ could be
neglected. Especially, the $\zeta^{LL}$ part moderates the
contradictions caused by $\xi^{LL}$ and $\xi^{LR}$. Furthermore, it
is interesting to note that our predictions for ${\cal B}(B\to\pi
K^{\ast}, \rho K)$, $A_{CP}(B\to\pi K^{\ast})$ and
$A_{CP}^{mix}(B^0\to\pi^0 K_S, \rho^0 K_S)$, listed in
Tables~\ref{tab_br}, \ref{tab_cp} and \ref{tab_mixcp}, respectively,
are all consistent with the experimental data within $2\sigma$.

\begin{figure}[t]
\begin{center}
\epsfxsize=15cm \centerline{\epsffile{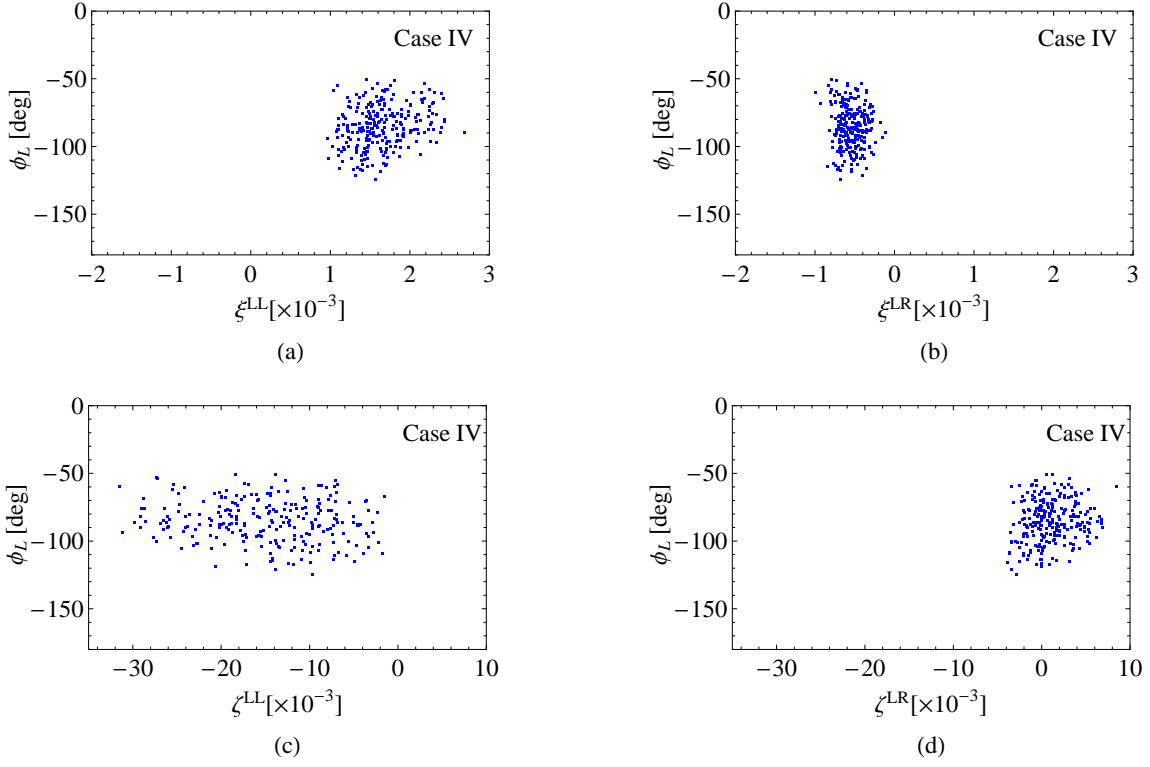}}
\centerline{\parbox{16cm}{\caption{\label{CaseIV}\small The allowed
regions for the parameters $\xi^{LL,LR}$, $\zeta^{LL,LR}$ and
$\phi_L$ in Case IV.}}}
\end{center}
\end{figure}

\subsubsection*{Case IV: Without any simplification of $B_{uu}^{L,R}$ and
$B_{dd}^{L,R}$, {\it i.e.,} arbitrary values of $\zeta^{LL,LR}$ and
$\xi^{LL,LR}$ are allowed.}

More generally, we give up any assumptions of the couplings
$B_{uu}^{L,R}$ and $B_{dd}^{L,R}$. Then, there are five arbitrary NP
parameters. As in Case III, we take ${\cal B}(B\to\pi K)$,
$A_{CP}(B\to\pi K)$ and $A_{CP}(B\to\rho K)$ as constraints and
present the predictions for the other observables.

The allowed regions for $\xi^{LL,LR}$, $\zeta^{LL,LR}$ and $\phi_L$
are shown in Fig.~\ref{CaseIV}, while the numerical results are
listed in the last column of Table~\ref{NPPareValue}. We find that,
similar to Case III, the values of $\xi^{LL,LR}$ are definitely
nonzero. The values of $\zeta^{LL}$ is a little larger than the one
in Case III, due to the interference effect caused by the parameter
$\zeta^{LR}$. Our predictions for ${\cal B}(B\to\pi K^{\ast}, \rho
K)$, $A_{CP}(B\to\pi K^{\ast})$ and $A_{CP}^{mix}(B^0\to\pi^0 K_S,
\rho^0 K_S)$, listed in Tables~\ref{tab_br}, \ref{tab_cp} and
\ref{tab_mixcp}, respectively, are consistent with the experimental
data within $2\sigma$.

\section{Conclusions}

Motivated by the recent observed large difference $\Delta A$ between
${\cal A}_{CP}(B^{\mp}\to\pi^0 K^{\mp})$ and $ A_{CP}(B^{0}\to
K^{\pm}\pi^{\mp})$, we have investigated the effect of family
non-universal $Z^{\prime}$ model and pursued possible solutions to
the observed ``$\pi K$ puzzle''. Moreover, we have also taken into
account the constraints from the $B\to \pi K^{\ast}$, $\rho K$
decays, which also involve the same quark level $b\to s \bar{q}q$
($q=u,d$) transitions. Our main conclusions are summarized as:

\begin{itemize}
\item The $Z^{\prime}$ contributions to the coefficients of operators $O_7$ and $O_9$ ($\xi^{LL}$ and
$\xi^{LR}$) with $\phi_L\sim-86^{\circ}$ are crucial to bridge the
discrepancy of $A_{CP}(B^-\to\pi^0 K^-)$ between theoretical
prediction and experimental data. However, they are definitely
unequal and opposite in sign.

\item The $Z^\prime$ contributions to the coefficients of QCD penguins operator $O_3$ related to
$\zeta_{LL}$ are required to moderate the contradiction of
$A_{CP}(B^-\to\rho^0 K^-)$ and $A_{CP}(B^0\to\pi^+ K^-)$ to thier
experimental values induced by $\xi^{LL}$ and $\xi^{LR}$,
respectively, even though they are helpless to resolve the observed
``$\pi K$ puzzle''. On the other hand, the $Z^\prime$ contributions
to $C_5(\zeta^{LR})$ are inessential.

\item For all of the four cases, a new weak phase associated with
the chiral $Z^\prime$ couplings, with a value about $-86^{\circ}$,
is always required for the ``$\pi K$ puzzle''.
\end{itemize}

Combing the up-to-date experimental measurements of $B\to \pi K$,
$pi K^{\ast}$ and $\rho K$ decays, the family non-universal
$Z^{\prime}$ model is found to be helpful to resolve the
observed``$\pi K$ puzzle''. It is also reminded that more refined
measurements of the mix-induced CP asymmetries in the $B^0\to \pi^0
K_S$ and $\rho^0 K_S$ decays are required to confirm or refute the
NP signals. In the following years, the precision of measurements
for these observables is expected to be much improved, which will
then shrink and reveal the $Z^\prime$ parameter spaces.

{\it 
Note added:  When the paper is finished, we are aware of the interesting paper by 
Barger et al.\cite{barger}.  Although our topics are very similar, we have taken into account of not
only the CP asymmetries but also the branching ratios of the correlated decay modes 
to constrain $Z^\prime$ couplings. Moreover, our approaches for the hadronic dynamics 
are different.}

\section*{Acknowledgments}
The work is supported by National Science Foundation under contract
Nos.10675039 and 10735080. X.~Q.~Li acknowledges support from the
Alexander-von-Humboldt Foundation.

\begin{appendix}

\section*{Appendix~A: decay amplitudes in the SM with QCDF}
The decay amplitudes for $B\to\pi K$ decays are recapitulated from
Ref.~\cite{Beneke3}
\begin{eqnarray}
{\cal A}_{B^-\to\pi^- \bar{K}}^{\rm SM}
   &=& \sum_{p=u,c}V_{pb}V_{ps}^{\ast} A_{\pi \bar{K}} \Big[
    \delta_{pu}\,\beta_2 + \alpha_4^p - \half\alpha_{4,{\rm
    EW}}^p +\beta_3^p+\beta_{3,{\rm
    EW}}^p\Big],
\label{amp1_SM}
\end{eqnarray}
\begin{eqnarray}
\sqrt2\, {\cal A}_{B^-\to\pi^0 K^-}^{\rm SM}
   &=& \sum_{p=u,c}V_{pb}V_{ps}^{\ast} \biggl\{ A_{\pi^0 K^-} \Big[
    \delta_{pu}\,(\alpha_1+\beta_2) + \alpha_4^p + \alpha_{4,{\rm
    EW}}^p +\beta_3^p+\beta_{3,{\rm EW}}^p\Big]\nonumber\\
   &&+  A_{ K^- \pi^0}\Big[\delta_{pu}\,\alpha_2+\frac{3}{2}
   \alpha_{3,{\rm EW}}^p\Big]\biggl\},
\label{amp2_SM}
\end{eqnarray}
\begin{eqnarray}
{\cal A}_{\bar{B}^0\to\pi^+ K^-}^{\rm SM}
   &=& \sum_{p=u,c}V_{pb}V_{ps}^{\ast} A_{\pi^+ K^-} \Big[
    \delta_{pu}\,\alpha_1 + \alpha_4^p + \alpha_{4,{\rm
    EW}}^p +\beta_3^p-\half\beta_{3,{\rm EW}}^p\Big],
\label{amp3_SM}
\end{eqnarray}
\begin{eqnarray}
\sqrt2\, {\cal A}_{\bar{B}^0\to\pi^0 \bar{K}^0}^{\rm SM}
   &=& \sum_{p=u,c}V_{pb}V_{ps}^{\ast} \biggl\{ A_{\pi^0 \bar{K}^0}
   \Big[
    -\alpha_4^p + \half\alpha_{4,{\rm
    EW}}^p -\beta_3^p+\half\beta_{3,{\rm EW}}^p\Big]\nonumber\\
   &&+  A_{ \bar{K}^0
   \pi^0}\Big[\delta_{pu}\,\alpha_2+\frac{3}{2}\alpha_{3,{\rm
   EW}}^p\Big]\biggl\}\,,
\label{amp4_SM}
\end{eqnarray}
where the explicit expressions for the coefficients
$\alpha_i^p\equiv\alpha_i^p(M_1M_2)$ and
$\beta_i^p\equiv\beta_i^p(M_1M_2)$ can also be found in
Ref.~\cite{Beneke3}. Note that expressions of the hard-spectator
terms $H_i$ appearing in $\alpha_i^p$ and the weak annihilation ones
appearing in $\beta_i^p$ should be replaced by our recalculated ones
listed in Appendix B. The decay amplitudes of $B\to\pi K^{\ast}$ and
$B\to\rho K$ decays could be obtained from the above results by
replacing $(\pi K)\to (\pi K^{\ast})$ and $(\pi K)\to (\rho K)$,
respectively.

\section*{Appendix~B: The hard-spectator and annihilation corrections
with the infrared finite gluon propagator}

With the infrared finite gluon propagator to cure the end-point
divergences, the hard-spectator corrections in $B\to PP$ and $PV$
decays can be expressed as~\cite{YDYang}
\begin{equation}
H_i(M_1M_2)= \frac{B_{M_1 M_2}}{A_{M_1 M_2}} \int_0^1dxdyd\xi
\frac{\alpha_s(q^2)}{\xi}\Phi_{B1}(\xi)\Phi_{M_2}(x)\Big[\frac{\Phi_{M_1}(y)}
{\bar{x}(\bar{y}+\omega^2(q^2)/\xi)}+r_\chi^{M_1}\frac{\phi_{m_1}(y)}
{x(\bar{y}+\omega^2(q^2)/\xi)}\Big],
\label{hard1}
\end{equation}
for the insertion of operators $Q_{i=1-4,9,10}$,
\begin{equation}
H_i(M_1M_2)= -\frac{B_{M_1 M_2}}{A_{M_1 M_2}} \int_0^1dxdyd\xi
\frac{\alpha_s(q^2)}{\xi}\Phi_{B1}(\xi)\Phi_{M_2}(x)\Big[\frac{\Phi_{M_1}(y)}
{x(\bar{y}+\omega^2(q^2)/\xi)}+r_\chi^{M_1}\frac{\phi_{m_1}(y)}{\bar{x}
(\bar{y}+\omega^2(q^2)/\xi)}\Big], \label{hard2}
\end{equation}
for $Q_{i=5,7}$, and $H_i(M_1M_2)=0$ for $Q_{i=6,8}$. When both
$M_1$ and $M_2$ are pseudoscalars, the final building blocks for
annihilation contributions can be expressed as~\cite{YDYang}
\begin{eqnarray}
A_1^i&=&\pi\int_0^1dxdy\alpha_s(q^2)\biggl\{\Big[
\frac{\bar{x}}{(\bar{x}y-\omega^2(q^2)+i\epsilon)(1-x\bar{y})}+
\frac{1}{(\bar{x}y-\omega^2(q^2)+i\epsilon)\bar{x}}\Big]\Phi_{M_1}(y)\Phi_{M_2}(x)
\nonumber\\
&&+\frac{2}{\bar{x}y-\omega^2(q^2)+i\epsilon}r_\chi^{M_1}r_\chi^{M_2}\phi_{m_1}
(y)\phi_{m_2}(x)\biggl\}~,\label{anni1}
\end{eqnarray}
\begin{eqnarray}
 A_2^i&=&\pi\int_0^1dxdy\alpha_s(q^2)\biggl\{\Big[
\frac{y}{(\bar{x}y-\omega^2(q^2)+i\epsilon)(1-x\bar{y})}+
\frac{1}{(\bar{x}y-\omega^2(q^2)+i\epsilon)y}\Big]\Phi_{M_1}(y)\Phi_{M_2}(x)
\nonumber\\
&&+\frac{2}{\bar{x}y-\omega^2(q^2)+i\epsilon}r_\chi^{M_1}r_\chi^{M_2}\phi_{m_1}(y)
\phi_{m_2}(x)\biggl\},~\label{anni2}
\end{eqnarray}
\begin{eqnarray}
A_3^i&=&\pi\int_0^1dxdy\alpha_s(q^2)\biggl\{\frac{2\bar{y}}{(\bar{x}y-\omega^2(q^2)
+i\epsilon)(1-x\bar{y})}
r_\chi^{M_1}\phi_{m_1}(y)\Phi_{M_2}(x)\nonumber\\
&&-\frac{2x}{(\bar{x}y-\omega^2(q^2)+i\epsilon)(1-x\bar{y})}r_\chi^{M_2}(x)
\phi_{m_2}(x)\Phi_{M_1}(y)\biggl\}~,\label{anni3}
\end{eqnarray}
\begin{eqnarray}
A_1^f&=&A_2^f=0,~\label{anni4}
\end{eqnarray}
\begin{eqnarray}
A_3^f&=&\pi\int_0^1dxdy\alpha_s(q^2)\biggl\{\frac{2(1+\bar{x})}{(\bar{x}y-
\omega^2(q^2)+i\epsilon)\bar{x}}
r_\chi^{M_1}\phi_{m_1}(y)\Phi_{M_2}(x)\nonumber\\
&&+\frac{2(1+y)}{(\bar{x}y-\omega^2(q^2)+i\epsilon)y}r_\chi^{M_2}(x)\phi_{m_2}(x)
\Phi_{M_1}(y)\biggl\}~.\label{anni5}
\end{eqnarray}
When $M_1$ is a vector meson and $M_2$ a pseudoscalar, the sign of
the second term in $A_1^i$, the first term in $A_2^i$, and the
second terms in $A_3^i$ and $A_3^f$ need to be changed. When $M_2$
is a vector meson and $M_1$ a pseudoscalar, one only has to change
the overall sign of $A_2^i$.

\section*{Appendix~C: Theoretical input parameters}

\subsection*{C1. CKM matrix elements}
For the CKM matrix elements, we adopt the Wolfenstein
parameterization~\cite{Wolfenstein:1983yz} and choose the four
parameters $A$, $\lambda$, $\rho$ and $\eta$
as~\cite{Charles:2004jd}
\begin{equation}
A=0.798^{+0.023}_{-0.017}, \quad
\lambda=0.22521^{+0.00083}_{-0.00082}, \quad
\overline{\rho}=0.141^{+0.035}_{-0.021}, \quad
\overline{\eta}=0.340\pm0.016,
\end{equation}
with $\overline{\rho}=\rho\,(1-\frac{\lambda^2}{2})$ and
$\bar{\eta}=\eta\,(1-\frac{\lambda^2}{2})$.

\subsection*{C2. Quark masses and lifetimes}
As for the quark masses, there are two different classes appearing
in our calculation. One type is the current quark mass which appears
in the factor $r_\chi^M$ through the equation of motion for quarks.
This type of quark masses is scale dependent and denoted by
$\overline{m}_q$. Here we take
\begin{equation}
\overline{m}_s(\mu)/\overline{m}_q(\mu)=27.4\pm0.4\,~\cite{HPQCD:2006},\quad
\overline{m}_{s}(2\,{\rm GeV}) =87\pm6\,{\rm
MeV}\,~\cite{HPQCD:2006}, \quad
\overline{m}_{b}(\overline{m}_{b})=4.20^{+0.17}_{-0.07}\,{\rm
GeV}~\cite{PDG08}\,,
\end{equation}
where $\overline{m}_q(\mu)=(\overline{m}_u+\overline{m}_d)(\mu)/2$,
and the difference between $u$ and $d$ quark is not distinguished.

The other one is the pole quark mass appearing in the evaluation of
penguin loop corrections, and denoted by $m_q$. In this paper, we
take~\cite{PDG08}
\begin{equation}
 m_u=m_d=m_s=0, \quad m_c=1.61^{+0.08}_{-0.12}\,{\rm GeV}, \quad m_b=4.79^{+0.19}_{-0.08}\,{\rm GeV}.
\end{equation}

As for the B-meson lifetimes, we take~\cite{PDG08} $\tau_{B_{u}} =
1.638\,{\rm ps}$ and $ \tau_{B_{d}}=1.530\,{\rm ps}$, respectively.

\subsection*{C3. The decay constants and form factors}
In this paper, we take the heavy-to-light transition form
factors~\cite{BallZwicky}
\begin{eqnarray}
 & &F^{B\to \pi}_{0}(0)=0.258\pm0.031, \quad
     F^{B\to {K}}_{0}(0)=0.331\pm0.041,\quad
     V^{B\to K^\ast}(0)=0.411\pm0.033,\nonumber\\
 & & A_0^{B\to K^\ast}(0)=0.374\pm0.034,\quad
     A_1^{B\to K^\ast}(0)=0.292\pm0.028,\quad
     V^{B\to \rho}(0)=0.323\pm0.030, \nonumber\\
  & &A_0^{B\to \rho}(0)=0.303\pm0.029,\quad
     A_1^{B\to \rho}(0)=0.242\pm0.023.
\end{eqnarray}
and the decay constants
\begin{eqnarray}
&&f_{B}=(216\pm22)~{\rm MeV}~\cite{Gray:2005ad}\,, \quad
    f_\pi=(130.4\pm0.2)~{\rm MeV}~\cite{PDG08}\,,\quad
    f_{K}=(155.5\pm0.8)~{\rm MeV}~\cite{PDG08}\,,\nonumber\\
&&f_{K^{\ast}}=(217\pm5)~{\rm MeV}~\cite{BallZwicky}, \quad
    f_{\rho}=(209\pm2)~{\rm MeV}~\cite{BallZwicky}.
\end{eqnarray}

\subsection*{C4. The LCDAs of mesons and light-cone projector operators.}
The light-cone projector operators of light mesons in momentum space
read~\cite{Terentev,Beneke3}
\begin{equation}\label{pimeson3}
   M_{\alpha\beta}^P = \frac{i f_P}{4} \left[
   \pslash\,\gamma_5\,\Phi_P(x) - \mu_P\gamma_5\,
   \frac{\kslash_2\,\kslash_1}{k_2\cdot k_1}\,\Phi_p(x)
   \right]_{\alpha\beta}\,,
\end{equation}
\begin{equation}\label{provect2}
   \left( M^V_\parallel \right)_{\alpha\beta}
   = -\frac{i f_V}{4} \left[ \pslash\,\Phi_V(x)
   - \frac{m_V f_V^\perp}{f_V}\,
   \frac{\kslash_2\,\kslash_1}{k_2\cdot k_1}\,\Phi_v(x)
   \right]_{\alpha\beta}\,,
\end{equation}
where $f_{P,V}$ are the decay constants, and $\mu_P=m_b r_\chi^P/2$,
with the chirally-enhanced factor $r_\chi^P$ defined as
\begin{eqnarray}
r_{\chi}^{\pi}(\mu)&=&\frac{2m_{\pi}^2}{m_b(\mu)2m_{q}(\mu)}\,,\quad
r_{\chi}^{K}(\mu)=\frac{2m_{K}^2}{m_b(\mu)(m_{q}+m_s)(\mu)}\,,
\end{eqnarray}
where the quark masses are all running masses defined in the
$\overline{\rm MS}$ scheme. For the LCDAs of mesons, we use their
asymptotic forms~\cite{projector,formfactor}
\begin{equation}
\Phi_{P,V}(x)=6\,x(1-x)\,, \quad \phi_p (x)=1\,, \quad \phi_v(x)=3(2
x-1)\,.
\end{equation}

As for the B-meson wave function, we take the form~\cite{YYPhiB}
\begin{equation}
\Phi_B(\xi)=N_B\xi(1-\xi)\textmd{exp}\Big[-\Big(\frac{M_B}{M_B-m_b}\Big)^2(\xi-\xi_B)^2\Big],
\end{equation}
where $\xi_B\equiv1-m_b/M_B$, and $N_B$ is the normalization
constant to insure that $\int_0^1 d\xi\Phi_B(\xi)=1$.

\end{appendix}


 \end{document}